\documentclass[aip,english,graphicx]{revtex4-1}

\usepackage{amsmath}
\usepackage{amsthm}
\usepackage{txfonts}
\usepackage{babel}
\usepackage{graphicx}

\theoremstyle{plain}

\newtheorem{theorem}{Theorem}[section]
\newtheorem{conjecture}[theorem]{Conjecture}

\newcommand{\RealVect}[1]{{\mathbb R}^{#1}}
\newcommand{\Hdrest}[1]{\Hd_{#1}}

\def\Rd{\RealVect{d}} 
\def\Rp{\RealVect{p}} 
\def\MA{{\mathcal M}(A)} 
\def\MAp{{\mathcal M}_1(A)}
\def\Es{{\mathcal E}_s} 
\def\Ed{{\mathcal E}_d} 
\def\Hd{{\mathcal H}^d} 
\def\Hdr{\Hdrest{A}} 

\DeclareMathOperator{\supp}{supp}

\begin{document}

\title{Observed Asymptotic Differences in Energies of Stable and
  Minimal Point Configurations on $\mathbb{S}^2$ and the Role of
  Defects}

\author{M. Calef}
\email{mcalef@lanl.gov}
\thanks{The work of M. Calef was performed under the auspices of the
  National Nuclear Security Administration of the US Department of
  Energy at Los Alamos National Laboratory under Contract
  No. DE-AC52-06NA25396 LA-UR-13-27573.}
\affiliation{Computational Physics and Methods, Los Alamos National
  Laboratory}

\author{W. Griffiths}
\email{whitney.griffiths@baml.com}
\affiliation{Structured Credit, Bank of America}

\author{A. Schulz} 
\email{alexia.schulz@ll.mit.edu}
\affiliation{Cyber Security and
  Information Sciences, MIT Lincoln Laboratory}

\author{C. Fichtl}
\email{cfichtl@lanl.gov}
\affiliation{Improvised and Foreign Designs, Los Alamos National
  Laboratory}

\author{D. Hardin}
\email{doug.hardin@vanderbilt.edu}
\thanks{The research of D. Hardin was supported, in part, by the
  U. S. National Science Foundation under grant DMS-1109266.}
\affiliation{Department of Mathematics, Vanderbilt University}

\date{\today}

\begin{abstract}
Configurations of $N$ points on the two-sphere that are stable with
respect to the Riesz $s$-energy have a structure that is largely
hexagonal. These stable configurations differ from the configurations
with the lowest reported $N$-point $s$-energy in the location and
structure of defects within this hexagonal structure.  These
differences in energy between the stable and minimal configuration
suggest that energy scale at which defects play a role. This work uses
numerical experiments to report this difference as a function of $N$,
allowing us to infer the energy scale at which defects play a
role. This work is presented in the context of established estimates
for the minimal $N$-point energy, and in particular we identify terms
in these estimates that likely reflect defect structure.
\end{abstract}

\pacs{ 89.75.Kd, 89.75.Da, 71.10.-w }

\keywords{Keywords: Riesz Energy, Thomson Problem, Defects}

\maketitle

\section{Introduction}

\begin{figure}
\begin{center}
\includegraphics[height=7cm]{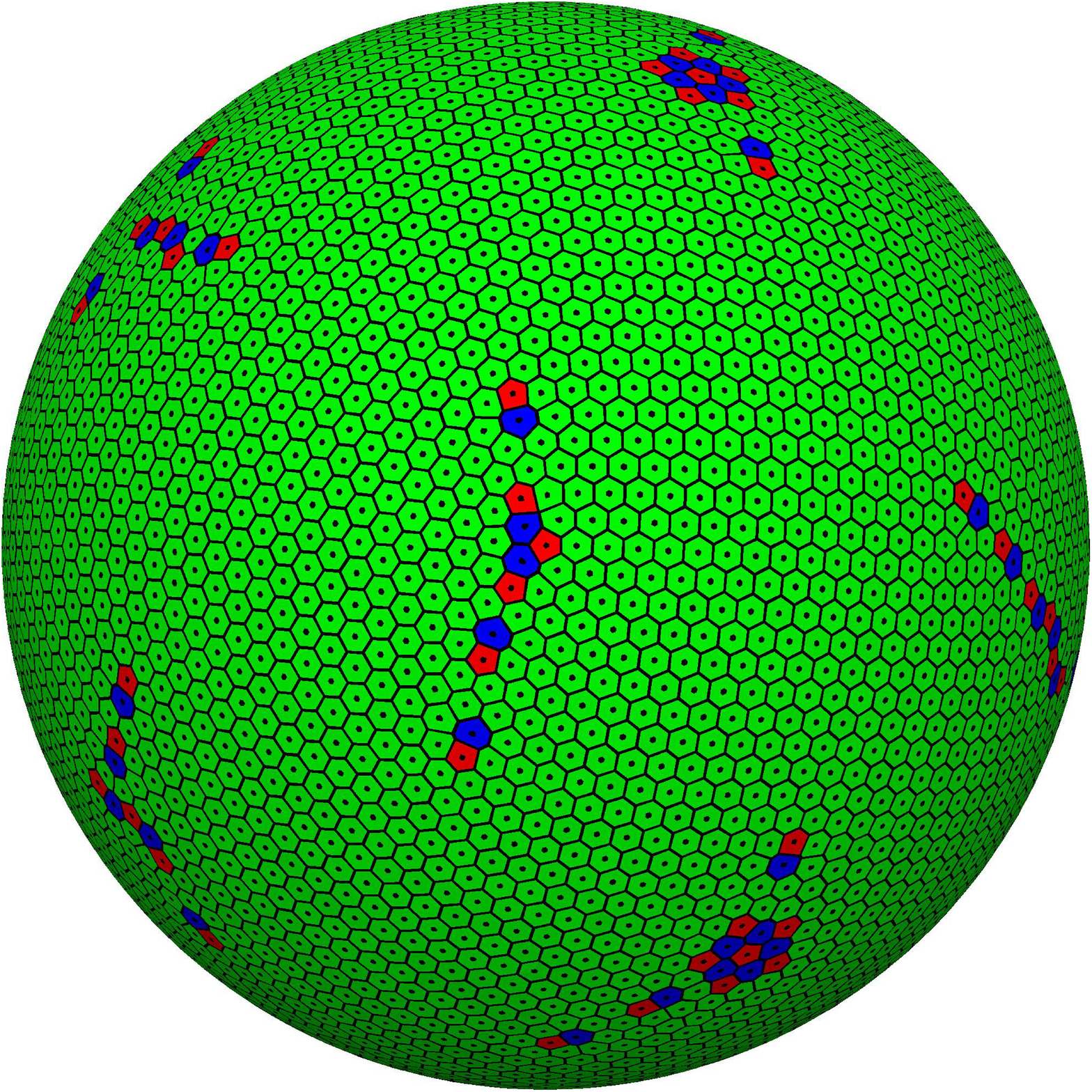}
\hspace{1cm}
\includegraphics[height=7cm]{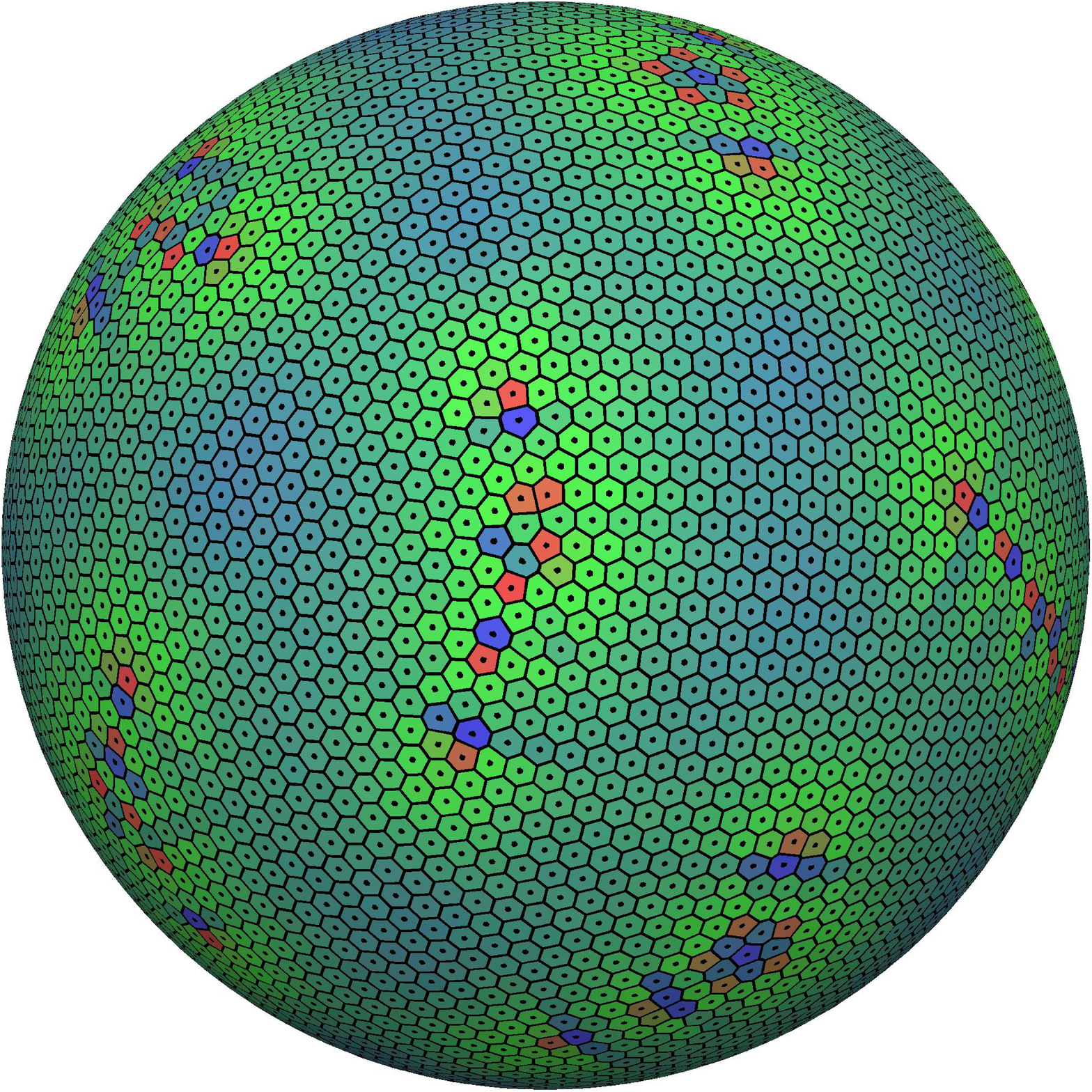}
\end{center}
\caption{\label{fig:Sample} The configuration with the lowest observed
  energy for $s=1$ and $N=4352$ points generated by Wales, McKay and
  Altschuler~\cite{WalesMcKayAltschuler:2009}.  Each point is depicted
  as a dot on the surface of the sphere surrounded by its Voronoi
  cell, which we computed with QHull~\cite{QHull:1996}.  In the image
  on the left the five sided cells are gray (red in the online
  version), the six sided cells are light gray (green in the online
  version) and the seven sided cells are dark gray (blue in the online
  version).  The image on the right shows the same configuration in
  the same orientation, but with the Voronoi cells colored by point
  energy.  (In the online version blue indicates the lowest point
  energies, green average point energies and red, the highest.)  Note
  that the fluctuations in point energy extend out from the defects
  into the surrounding ``hexagonal sea''.}
\end{figure}

The famous Thomson Problem~\cite{Thomson1} is to find, for an
arbitrary natural number $N$, a configuration of $N$ classical
electrons on the unit sphere, $\mathbb{S}^2$, that minimizes the
Coulomb energy.  There is no general theoretical solution to this
problem.  The apparent obstacle is strong evidence suggesting that
the ground state for the Coulomb potential in two dimensions has a
hexagonal structure.  The sphere, however, cannot be tiled exclusively
with hexagons. If one places points numbered $i=1,\ldots,N$ on the
sphere, and divides the sphere into Voronoi cells centered at each of
the $N$ points, then the Euler characteristic of the sphere ensures
that
$$
\sum_{i=1}^N 6 - V_i  = 12,
$$ where $V_i$ is the number of sides of the Voronoi cell associated
with the $i^\text{th}$ point.  One can see examples of these
non-hexagonal Voronoi cells, which are commonly referred to as defects
or scars, in Figure~\ref{fig:Sample}.  Finding the energy minimizing
configuration will likely require finding the right defect structure.
Many numerical techniques that aim to identify minimal energy
configurations rely on gradient information and tend to find
configurations that are stable, but not minimal.  These stable
configurations also have a local hexagonal structure, but differ from
one another (and presumably the energy minimizing configuration)
largely in location and structure of defects.  

A natural question to ask is: how much does the energy change as the
structure and location of defects changes? Because stable configurations
differ from the minimal configuration in location and structure of
defects, a related question is: how much does the average energy of
stable configurations differ from the true minimal energy? We answer
this empirically by developing a large library of stable
configurations and comparing the resulting average energy of the
stable configurations with the lowest observed energy.

Minimal energy is often approximated in an asymptotic expansion, in
$N$, and we compare the difference between the average and lowest
observed energy with the terms in these asymptotic expansions.  That
is, we empirically identify the terms in the asymptotic expansion that
approximate the lowest observed energy, but not the average energy.
We believe that these terms likely reflect characteristics of defects.

This work has value in several ways.  First, because the energy of any
configuration of points on the sphere is an upper bound for the
minimal energy, these results provide a lower bound for the
difference between the average energy of stable configurations and the
minimal energy.  Second, there are methods that have a controllable
error bound for quickly approximating the pairwise energy, most
notably the Fast Multipole Method~\cite{GreengardRokhlin:1987}.  For
such approximations the results in this paper will help select the
error bound necessary to distinguish stable configurations from
minimal configurations.  Finally, this work suggests which terms in
the asymptotic expansion will require an understanding of defect
structure.

The rest of the paper is organized as follows:
Section~\ref{sec:background} review some of the relevant work.
Section~\ref{sec:genconf} describes our method for generating stable
configurations, and reports properties of these stable configurations.
Section~\ref{sec:minest} compares theory and conjecture with minimal
observed energy and reports the observed asymptotic differences
between the average energy of stable configurations and minimal
observed energy.  Additionally, we examine and extend some conjectures
regarding the second order term for the Thomson problem.  In
Section~\ref{sec:conclusion} we summarize our results.

\section{Background}\label{sec:background}

Some of the earliest computational work on the Thomson Problem was
done by Erber and Hockney~\cite{ErbHock2,ErbHock1} where they rely on
optimization techniques to search for minimal energy configurations.
Rakhmanov, Saff and Zhou~\cite{RakSaffZhou1} presented a comprehensive
search for the minimal energies for $N$ up to $200$ for the
logarithmic as well as Coulomb energies.  Morris, Deavon and
Ho~\cite{MorDevHo1} used a genetic algorithm in an effort to avoid
becoming trapped in stable non-minimal configurations.  An important
effort to constructively generate candidate minimal energy
configurations came from Altschuler, Williams, Ratner, Tipton, Stong,
Dowla and Wooten~\cite{AWRTSDW:1997}, where the authors of that paper
identified configurations with twelve point defects and high symmetry.
These configurations were later shown not to be minimal by
P\'erez-Garrido, Dodgson, Moore, Ortuno and
Diaz-Sanchez~\cite{PerezGarrido:1997} and P\'erez-Garrido, Dodgson,
Moore~\cite{PerezGarrido:1997a}.  These authors found that, as $N$
increased, the defects were not point defects, but had considerable
structure such as those in Figure~\ref{fig:Sample}.  Efforts to
understand and characterize this structure, as well as find minimal
energy configurations, include the work of Wales and
Ulker~\cite{WalesUlker:2006} and Wales, McKay and
Altschuler~\cite{WalesMcKayAltschuler:2009}.  The results of the
experiments described in these two publications are collected in the
Cambridge Cluster Database~\cite{CCD1}\,\cite{CCD2}, and provide, to
our knowledge, the lowest observed energies for the Thomson Problem.
Bowick, Cacciuto, Nelson and Travesset~\cite{BCNT1} use a continuum
elasticity model to describe the interaction of defects.  In these
works the empirical evidence is that configurations with low energy
consist of a ``hexagonal sea'' with complex defects at the vertices of
an icosahedron inscribed in $\mathbb{S}^2$.

Theoretical examinations of the Thomson Problem provide valuable
insights and language for the problem, and we review some of the
relevant theory here.  Let $\omega_N$ denote a set $\{{\bf
  x}_1,\ldots,{\bf x}_N\}$ of $N$ distinct points in $\Rp$.  We
consider the following discrete energy of $\omega_N$
\begin{equation}\label{eq:energysum}
E_s(\omega_N) := \sum_{i=1}^N\sum_{j=1\,j\ne i}^N k_s(|{\bf x}_i -
{\bf x}_j|),
\end{equation}
where $k_s$ is the function given by
$$
k_s(r)=\left\{
\begin{array}{cc}
r^{-s} & \text{for $s > 0$} \\
-\log r & \text{for $s = 0$},
\end{array}
\right.
$$ and where $|\cdot|$ is the Euclidean norm inherited from $\Rp$.
Note that many papers on this topic report an energy where the second
sum is over $j=i+1,\ldots,N$ leading to a factor of two difference in
our values for energy.  The functions, $k_s$, are the Riesz
potentials, which are a natural generalization of the Coulomb
potential. The questions in which we are interested apply to Riesz
potentials in general, and we present results for the Riesz potentials
corresponding to $s=0$, $1$, $2$, and $3$.  We denote the point
($s$-)energy of the $i^\text{th}$ point in $\omega_N$ by
$$
U_s^{i, \omega_N} := \sum_{j=1\,j\ne i}^N k_s(|{\bf x}_i - {\bf x}_j|)
\quad\text{and then the total energy is given by}\quad
E_s(\omega_N) = \sum_{i=1}^NU_s^{i, \omega_N}.
$$ For any compact set $A \subset \Rp$ of Hausdorff dimension $d>0$,
the lower semi-continuity of $k_s$ ensures that there is at least one
configuration contained in $A$, which we denote $\omega_N^{s,A}$, that
satisfies
$$
E_s(\omega_N^{s,A}) = \Es(A,N) := \inf \{ E_s(\omega_N) : \omega_N
\subset A\,\text{and}\, {\bf x}_i \ne {\bf x}_j\, \text{for all $i \ne j$}\}.
$$ That is to say, there is at least one energy-minimizing
configuration, $\omega_N^{s,A}$, and the minimal $N$-point $s$-energy
is denoted ${\mathcal E}_s(A,N)$. In this setting one can search for
an expansion of the minimal energy as a function of $N$ of the form
\begin{equation}\label{eq:expansion}
{\mathcal E}_s(A,N) \approx C_1 N^{\alpha_1} + C_2 N^{\alpha_2} +
\ldots .
\end{equation}  In certain cases, e.g. $s=0$ and $s=d$, this expansion will also
include logarithmic terms.

In the general case where $A$ is any $d$ dimensional compact set and
$s<d$, P\'olya and Szeg\"o establish the first order
term~\cite{PolyaSzego:1931} by connecting the asymptotic behavior of
the discrete minimal energy with a continuum problem.  Specifically,
let $\MA$ denote the positive Borel measures supported on $A$, and
$\MAp\subset\MA$ denote the Borel probability measures supported on
$A$.  One may interpret $\mu\in \MA$ as a continuous charge
distribution and consider the energy functional defined for any $\mu
\in \MA$, by
$$
I_s(\mu) := \iint k_s(|{\bf x}-{\bf y}|)\, d\mu ({\bf y})d\mu ({\bf x}).
$$ Analogous to the discrete point energy, $U_s^{i,\omega_N}$, the
potential due to $\mu$ at a point ${\bf x}$, is
$$
U_s^{\mu}({\bf x}) := \int k_s(|{\bf x}-{\bf y}|)\, d\mu ({\bf y}),
\quad\text{and then}\quad
I_s(\mu) := \int U_s^{\mu}({\bf x})\, d\mu({\bf x}).
$$ 
There is a unique energy-minimizing measure $\mu^{s,A}\in\MAp$ so that
$$
I_s(\mu^{s,A}) < I_s(\mu)\quad \text{for all}\quad \mu \in \MAp 
\backslash \{ \mu^{s,A}\}.
$$ (cf.~\cite[pp. 131-133]{Landkof1} also G\"otz~\cite{Gotz1}
provides a proof of a key step without using standard Fourier
techniques.)  Further,
\begin{equation}\label{eq:constPot}
U_s^{\mu^{s,A}}({\bf x}) = I_s(\mu^{s,A})
\end{equation} for all ${\bf x} \in \supp \mu^{s,A}$ with the possible exception
of a set that supports no measures of finite energy (cf.~\cite[Theorem
  2.4]{Fuglede1}).  Roughly speaking Equation~\eqref{eq:constPot} asserts that
the potential is constant in regions where there is charge.  The
essence of the proof is that, if this were not the case, energy could
be decreased by moving charge from regions of high potential to
regions of low potential. 

The celebrated transfinite diameter result of P\'olya and Szeg\"o
relates the continuous and discrete problems as follows (also
cf.~\cite[pp. 160-162]{Landkof1}): for any
continuous function $f:A\to \mathbb{R}$ and any sequence of
energy-minimizing configurations $\{\omega_N^{s,A}\}_{N=2}^\infty$,
$$
\lim_{N\to\infty} \frac 1 N \sum_{ {\bf x} \in  \omega_N^{s,A}} f({\bf
  x})
= \int f\, d\mu^{s,A},
$$
and
\begin{equation}\label{eq:TFD}
\lim_{N\to\infty} \frac {\Es(A,N)}{N^2} = I_s(\mu^{s,A}).
\end{equation} For this range of $s$ the discrete minimal energy
configurations are converging in the weak-star topology of measures to
$\mu^{s,A}$.  The minimal energy grows as $N^2$, where the coefficient
is given by $I_s(\mu^{s,A})$.  The proof of these results indicates
that the first order approximation of the minimal energy is determined
by the global distribution of points within energy minimizing
configurations.  Kuijlaars and Saff have shown~\cite{KuijlaarsSaff1}
that the second order term on the sphere in the
expansion~\eqref{eq:expansion} grows as $N^{3/2}$ and the, still to be
proven, coefficient is conjectured to depend on the presumed local
hexagonal structure.

If $s\ge d$, then $I_s(\mu) = \infty$ for all $\mu \in \MA \backslash
\{0\}$, (cf.~\cite[Ch. 8]{Mattila1}) and other techniques are required
to estimate growth in minimal energy.  Hardin and
Saff~\cite{HardinSaff1} and Borodachov, Hardin and Saff~\cite{BHS1}
show that when $A$ has certain smoothness properties
$$
\lim_{N\to\infty} \frac 1 N \sum_{ {\bf x} \in
\omega_N^{s,A}} f({\bf x})
=
\frac 1 {|\Hdr|}  \int f\, d\Hdr, 
$$
and 
$$
\lim_{N\to\infty} \frac{\Es(A,N)}{N^{1+s/d}} =
\frac{C_{s,d}}{\Hd(A)^{s/d}}
\quad \text{for $s >d$, and}\quad
\lim_{N\to\infty} \frac{\Ed(A,N)}{N^{2}\log N} =
\frac{\Hd({\mathcal B}^d)}{\Hd(A)},
$$ where $\Hdr$ is the $d$ dimensional Hausdorff measure restricted to
$A$, $C_{s,d}$ is a constant that depends only on $d$ and $s$ and not
the underlying set $A$, and ${\mathcal B}^d$ is the closed unit ball
in $\Rd$.  These results demonstrate that for $s\ge d$ the asymptotic
distribution of points in energy-minimizing configurations is
uniform. Furthermore, the minimal $N$-point energy grows at a rate
exceeding $N^2$ and is determined largely by the local structure of
the energy minimizing configurations.  Indeed for the $d=2$ case,
numerical evidence supports the conjecture that $C_{s,d}$ is given by
a hexagonal zeta function evaluated at $s$, i.e. the sum of the
reciprocal non-zero distances in the hexagonal lattice raised to the
power $s$.  Brauchart, Hardin and Saff present a summary of theory and
conjecture regarding minimal energy configurations on the
sphere~\cite{BrauchartHardinSaff:2011}.

\section{Numerical Methods}\label{sec:genconf}

\subsection{Generating Candidate Minimal Energy Configurations}

To generate candidate configurations we begin with a random,
well-separated, initial configuration of points on $\mathbb{S}^2$ and
alternate between the Polak-Ribi\`ere variant of Conjugate Gradient
(cf.~\cite{NumRecC}) with a line minimization of the energy, and an
exact Newton's Method to find a root of the gradient.  To solve the
linear system arising in Newton's Method we use LAPACK~\cite{LAPACK}.

We use a direct evaluation of the energy sum, given in
Equation~\eqref{eq:energysum} omitting obvious duplicate calculations,
which involves $\mathcal{O}(N^2)$ terms, the smallest of which is
$k_s(2)$, while $\Es(\mathbb{S}^2,N)$ can grow into the hundreds of
millions for some values of $s$ and $N$ considered.  To control the
numerical round-off error associated with adding two numbers whose
ratio is far from unity (cf.~\cite{Higham:1993} for relevant work on
this problem) we logarithmically bin our summands.  By only adding
summands in the same bin, we bound the ratio of any two intermediate
summands to be added.  The final sum is computed by iterating over our
bins in increasing magnitude and summing their contents.

For $N=20,\ldots,180$ we ran thousands of trials.  For
$N=181,\ldots,500, 4352$ we ran tens to hundreds of trials.  We report
lowest observed energies on the sphere only for those $N$ where the
Cambridge Cluster Database provides a configuration with which we can
initialize our solver.

\subsection{Generating Stable Configurations}

The above optimization process leads to a candidate configuration
$\omega_N$, which we assume is close enough to a true stable
configuration $\bar \omega_N$ so that the linear approximation about
$\omega_N$ for the gradient
$$
0 = \nabla E_s(\bar \omega_N) \approx
\nabla E_s(\omega_N) + \nabla^2 E_s(\omega_N) (\bar \omega_N - 
\omega_N) 
$$ is reasonable. Here $\nabla E_s$ is the gradient of the energy with
respect to the free parameters that define $\omega_N$ and
$\nabla^2E_s$ is the Hessian represented in the same coordinates.
Were the Hessian invertible this would lead to the bound
$$ \frac {\| \nabla E_s(\omega_N) \|_2 }{\lambda_\text{min}} =
\|\nabla^2 E_s(\omega_N)^{-1}\|\, \| \nabla E_s(\omega_N) \|_2 \ge \|
\bar \omega_N - \omega_N \|_2 \ge \| \bar \omega_N - \omega_N
\|_\infty,
$$ where $\lambda_\text{min}$ is the smallest eigenvalue of the
Hessian, $\|\cdot\|_2$ is the unnormalized two-norm of the parameters
defining the argument, and $\| \cdot \|$ is the associated operator
two-norm.  Our choice of coordinates leads to three degrees of freedom
corresponding to rigid motions of the sphere and so the smallest three
eigenvalues of the Hessian are zero.  We assume a rotation and
reflection of $\bar \omega_N$ so that the difference between $\bar
\omega_N$ and $\omega_N$ and does not reflect these rigid motions.  We
let $\lambda^*_\text{min}$ denote the fourth lowest eigenvalue, then
we have the bound
$$
\frac {\| \nabla E_s(\omega_N) \|_2 }{\lambda^*_\text{min}}
\ge 
\| \bar \omega_N - \omega_N \|_\infty.
$$ We desire that 
$$
\| \bar \omega_N - \omega_N \|_\infty
\le 
\frac{\min_{i\ne j\in 1,\ldots,N} | {\bf x}_i - {\bf x}_j|}{10,000}
$$ Our reasoning is that the free parameters are the polar and
azimuthal angles, and, on the unit sphere, changes in position are
always bounded from above by changes in angle.  The above bound will
ensure that no point in $\omega_N$ is further from its corresponding
point in the true stable state by more than the arbitrary bound of one
ten-thousandth of the minimum separation in $\omega_N$. This is
ensured if
\begin{equation}\label{eq:StabCrit}
\frac {\| \nabla E_s(\omega_N) \|_2 }{\lambda^*_\text{min}}
\le 
\frac{\min_{i\ne j\in 1,\ldots,N} | {\bf x}_i - {\bf x}_j|}{10,000},
\end{equation}
where, again, we used LAPACK to compute $\lambda^*_\text{min}$.  We
reiterate that these estimates hinge on the assumption that the
gradient at the true stable state is well approximated by a linear
expansion of the gradient about the observed state. We keep candidate
configurations if Equation~\eqref{eq:StabCrit} holds or if the
configuration possesses the lowest observed energy.  

Note that Equation~\eqref{eq:StabCrit} is quite stringent.  As $N$
increases, the minimum pairwise separation between points goes as
$N^{-1/2}$.  In addition we have bounded from above the infinity-norm
with the unnormalized two-norm.  Such a bound is tight only when all
the components but one are zero.  This condition was relaxed for
$N=4352$, where we simply required that all but lowest three
eigenvalues be positive.

\subsection{Properties of Stable Configurations}

\begin{figure}
\includegraphics[height=9cm]{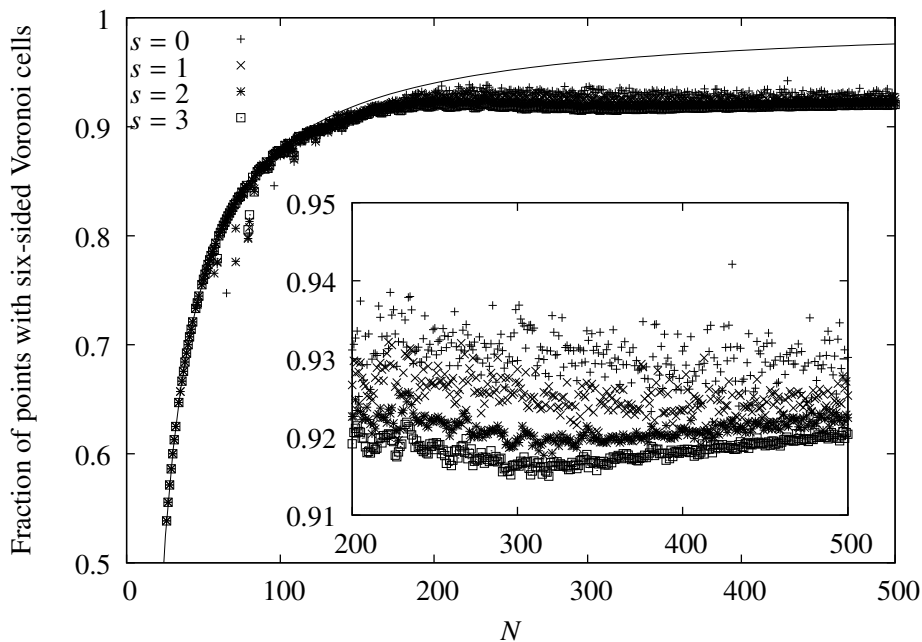}
\caption{\label{fig:frac_avg} This plot shows the fraction of points
  that have six-sided Voronoi cells for $N$ and for the values of $s$
  in which we are interested.  The inset plot provides more detail for
  $N=200,\ldots,500$.  In addition we've plotted the upper bound for
  this fraction assuming that no Voronoi cell has fewer than five
  sides.  Each data point is averaged over all the configurations,
  weighted by number of occurrences, for the specified $N$ and $s$.}
\end{figure}

In Figure~\ref{fig:frac_avg} one can see the average fraction of
points that have six-sided Voronoi cells.  For each $N$ and $s$, these
data are obtained by computing this fraction per configuration, and
then averaging over all the observed configurations and weighting by
the number of times the configuration occurred.  This is the same
averaging method we use when computing the average energy of stable
configurations.  As one can see this average fraction is better than $91$
percent for $N\ge 200$, supporting the claim that stable
configurations are largely hexagonal.

\begin{figure}
\includegraphics[height=9cm]{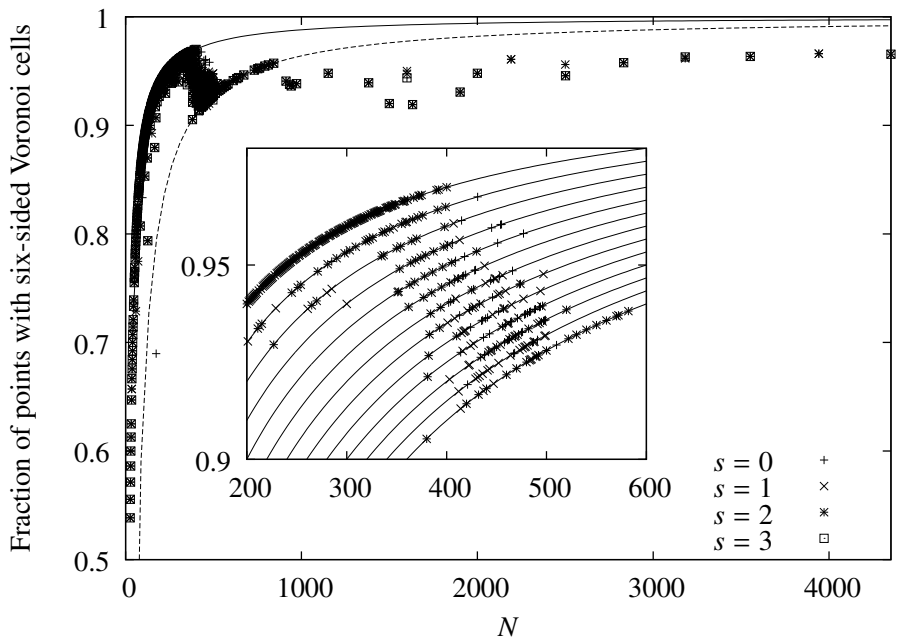}
\caption{\label{fig:frac_min} As in Figure~\ref{fig:frac_avg} we've
  plotted the fraction of points that have six-sided Voronoi cells.
  Here each data point corresponds to the configuration with the
  lowest observed energy.  In the outer plot the solid line is the
  upper bound on this fraction, while the dashed line indicates what
  this fraction would be if there were $12$ defects each consisting of
  a three points with five, seven and five sided Voronoi cells
  respectively.  In the inset one sees lines corresponding to
  $12,14,16,\ldots,36$ non six-sided Voronoi cells.}
\end{figure}

As a point of comparison, we've also computed this fraction for the
configurations that have the lowest observed energy.  This is shown in
Figure~\ref{fig:frac_min}.  One important feature of this plot is that
the configurations with the lowest observed energy have far more
non-six-sided Voronoi cells than the minimum allowed if no Voronoi
cell has fewer than five sides.  If one further assumes that no Voronoi
cell has more than seven sides, then the number of Voronoi cells with
other than six sides must be even.  This corroborates previous
observation that as $N$ increases, the defects cease to be single
points and develop structure.

\section{Asymptotics of Minimal Energy and Average Energy of Stable Configurations}\label{sec:minest}

In this section we compare theory and conjecture for the minimal
$N$-point energy with the lowest observed $N$-point energy.  In the
case $s=1$ we extend a conjecture for the second order term on
$\mathbb{S}^2$ to certain smooth manifolds.  We report the asymptotics
of the difference between the average and minimal observed energies
and compare this difference with terms in the asymptotic expansion.

Like all computational works of this type, we have no assurances that
the lowest available energies are indeed minimal.  Systematic errors
of this type would cause us to underestimate the difference between
the average and the minimal energies.  Consequently our results that
indicate that a term in the asymptotic expansion does not describe
both the average and minimal energy should be trusted more than
results indicate that a term does describe both the average and
minimal energies.

We shall use the following notation: $\tilde{\mathcal E}_s(A,N)$ is
the lowest \emph{observed} minimal $N$-point $s$-energy on a set $A$.
$R_s^n(A,N)$ is the difference between the minimal $N$-point
$s$-energy on $A$ and an $n$-term asymptotic expansion of the minimal
$s$-energy, while $\tilde R_s^n(A,N)$ is the difference between the lowest
\emph{observed} energy and the $n$-term expansion.

\subsection{The $s=1$ Case}\label{ss:s1}

This is the Thomson Problem, and the leading order term in the
asymptotic expansion of the minimal energy follows from the
transfinite diameter result in Equation~\eqref{eq:TFD}, i.e. for a set
$A$ of dimension $d>1$ it is $I_1(\mu^{1,A})N^2$. For the sphere a
simple calculation shows that $I_1(\mu^{1,\mathbb{S}^2}) = 1$. We now
review an existing conjecture for the second order term on
$\mathbb{S}^2$, and show how it may be generalized for compact
$2$-manifold $A$.  A trivial representation of the first order term
and the correction for a set $A$ is
\begin{equation}\label{eq:error1}
{\mathcal E}_1(A,N) = I_1(\mu^{1,A})N^2 + \sum_{i=1}^N\left( \sum_{j\ne i} \frac
1 {|{\bf x}_i - {\bf x}_j|^s} - I_1(\mu^{1,A})N \right).
\end{equation}
We shall consider the case that $\mu^{1,A}$ is absolutely continuous
with respect to $\Hdr$, the support of $\mu^{1,A}$ is all of $A$, and
Equation~\eqref{eq:constPot} holds for the entire support of
$\mu^{1,A}$, that is $U_s^{\mu^{1,A}}({\bf x}) = I_s(\mu^{s,A})$ for
all ${\bf x} \in A$.  These assumptions are satisfied for
$A=\mathbb{S}^2$.

The potential $U_s^{\mu}$ is linear in $\mu$ and so, with our
assumptions, we may write Equation~\eqref{eq:error1} as
\begin{equation}\label{eq:err-est1}
{\mathcal E}_1({A},N) = I(\mu^{1,A})N^2 + \sum_{i=1}^N\left( \sum_{j\ne i} \frac
1 {|{\bf x}_i - {\bf x}_j|} - U_1^{N\mu^{1,A}}({\bf x}_i)\right).
\end{equation} 

The above equation is exact regardless of where on $A$ we choose to
evaluate the potential $U_1^{N\mu^{1,A}}$.  However, choosing to
evaluate the potential at the points that form a minimal $N$-point
configuration suggests one way to express the correction: the point
energy for ${\bf x}_i$ should be corrected by subtracting the
potential at ${\bf x}_i$ due to $N$ times the equilibrium measure and
adding the energy due to the presence of the $N-1$ other discrete
points. In broader terms the point at ${\bf x}_i$ sees other discrete
points, not a smoothed out average density.

For the $i^\text{th}$ point, the correction given by
Equation~\eqref{eq:err-est1} may be written as two terms, which we
refer to as ``near'' and ``far'' contributions.
\begin{eqnarray}\label{eq:near-far}
\sum_{j\ne i} \frac 1 {|{\bf x}_i - {\bf x}_j|} - U_1^{N\mu^{1,A}}({\bf x}_i) &= 
\displaystyle{\left(\sum_{j\ne i} \frac {\exp ( -|{\bf x}_i - {\bf x}_j| / R)} {|{\bf x}_i - {\bf x}_j|} - 
\int \frac {\exp ( -|{\bf x}_i - {\bf y}| / R)} {|{\bf x}_i - {\bf y}|} d\mu^{1,A}({\bf y}) \right)} \\
\nonumber
&+
\displaystyle{\left(\sum_{j\ne i} \frac {1 - \exp ( -|{\bf x}_i - {\bf x}_j| / R)} {|{\bf x}_i - {\bf x}_j|} - 
\int \frac {1 - \exp ( -|{\bf x}_i - {\bf y}| / R)} {|{\bf x}_i - {\bf y}|} d\mu^{1,A}({\bf y}) \right) }
\end{eqnarray}

This decomposition is motivated by the reasoning presented by
Kuijlaars and Saff~\cite[Section 2]{KuijlaarsSaff1}, namely that
the second order correction for $0<s<2$ is determined by the local
structure.  Where Kuijlaars and Saff use a cutoff at radius $R$, we
use an exponential damping that allows use of the Poisson Summation
Formula and Ewald type arguments for the $s=1$ case.

We fix $R>0$ small enough so that $d\mu^{s,A}/d\Hdr$ changes on a
scale much larger than $R$, and we consider $N$ large enough so that
the nearest neighbor distance is much smaller than $R$. Then for most
$i$ we can expect a local hexagonal structure around ${\bf x}_i$ and
so we consider the following estimate for the near term in
Equation~\eqref{eq:near-far}:
$$
N^{-1/2}\left(\sum_{j\ne i} \frac {\exp ( -|{\bf x}_i - {\bf x}_j| / R)} {|{\bf x}_i - {\bf x}_j|} - 
\int \frac {\exp ( -|{\bf x}_i - {\bf y}| / R)} {|{\bf x}_i - {\bf
    y}|} d\mu^{1,A}({\bf y}) \right) 
$$
\begin{equation}\label{eq:approx1}
\approx  N^{-1/2} \left(
\sum_{{\bf x} \in D_i N^{-1/2}\Lambda\backslash\{0\}} 
\frac {\exp ( -|{\bf x}|/R )} {|{\bf x}|} 
- \frac 1 {|D_i N^{-1/2}\Lambda|}\int_{\RealVect{2}} 
\frac {\exp ( -|{\bf x}|/R )} {|{\bf x}|} 
d^2{\bf x}
\right)
\end{equation}
Here $\Lambda := \{ m{\bf r}_1 + n{\bf r}_2 : {\bf r}_1 = (1,0) ,
\,{\bf r}_1 = ( 1/2 , \sqrt{3}/2 ) \,\text{and}\, m,n \in \mathbb{Z}
\}$ is the hexagonal lattice of unit spacing, $D_i
\Lambda$ is the hexagonal lattice where the generating vectors have
been scaled by $D_i$, and $|D_i\Lambda| = \sqrt{3}D_i^2/2$ is
the area of the fundamental cell of the scaled lattice.  Finally,
$d^2{\bf x}$ denotes integration with respect to area. The essential
statement of the approximation in Equation~\eqref{eq:approx1} is that,
for most points in a configuration with low energy, the energy due to
neighboring points is well approximated by the energy due to the
neighboring points in an appropriately scaled hexagonal lattice, and
that the density represented by equilibrium measure changes little on
the scale of nearest neighbor separation.  This assumption is qualitatively
supported by Figure~\ref{fig:Sample} where most points are surrounded
by a local hexagonal structure.

We compute the sum over a lattice that is scaled by $D_iN^{-1/2}$,
which is intended to reflect the local point density of the energy
minimizing configuration near the point ${\bf x}_i$.  For the case $A
= \mathbb{S}^2$, $D_i$ is independent of $i$.  To generalize to an
arbitrary 2-manifold one may estimate $D_i$ as follows: Let $r$ be the
nearest-neighbor spacing. Assume that for large $N$, hence small $r$,
the Voronoi cells within $B({\bf x}_i , r_0)$ are all hexagonal and of
the same size.  This gives
\begin{equation}\label{eq:est1}
\#\left(\omega_N^{s,A}\cap B({\bf x}_i , r_0)\right) H_{r/2} \approx {\mathcal H}^2_A ( B( {\bf x}_i , r_0 ) ).
\end{equation} Here $\#$ indicates the number of points in the
following set. $H_{r/2}$ is the area of a hexagon of inner radius
$r/2$, which is $\sqrt{3} r^2/2$.  

The second estimate follows from the weak-star convergence of the
discrete minimal energy points to the equilibrium measure and the
assumption that $A \cap B({\bf x}_i,r_0)$ is $\mu^{1,A}$-almost clopen.
Then, for $N$ sufficiently high,
\begin{equation}\label{eq:est2}
\frac {\#\left(\omega_N^{s,A}\cap B({\bf x}_i , r_0)\right)} N \approx
\mu^{1,A} ( B({\bf x}_i , r_0 ) ).
\end{equation}

Dividing \eqref{eq:est2} by \eqref{eq:est1} gives, for $N$
sufficiently large
$$
\frac 2 {\sqrt{3} r^2 N} = 
\frac {\mu^{1,A}(B({\bf x}_i , r_0))}
{ {\mathcal H}^2_A ( B ({\bf x}_i, r_0) )}.
$$ As $r_0$ decreases to zero, the right hand side tends toward the
Radon-Nikod\'ym derivative of $\mu^{1,A}$ with respect to ${\mathcal
  H}^2_A$ and we have that the nearest neighbor spacing $r$, and the
appropriate scaling for the lattice at ${\bf x}_i$, is given by
$$
r = \sqrt{
\frac 2 {\sqrt{3} N} \left( 
\frac {d\mu^{1,A}}{d{\mathcal H}^2_A}
({\bf x}_i)\right)^{-1}
}\qquad\text{hence}\qquad
D_i = 
\sqrt{
\frac 2 {\sqrt{3}} \left( 
\frac {d\mu^{1,A}}{d{\mathcal H}^2_A}
({\bf x}_i)\right)^{-1}
}
$$
With some substitutions, the limit as $N$ grows to infinity of \eqref{eq:approx1} may be
expressed as 
\begin{equation}\label{eq:to-evaluate}
\frac 1 { D_i } \lim_{R\to \infty} \left(
\sum_{{\bf x} \in \Lambda \backslash \{0 \}} \frac {\exp(-|{\bf x}|/R)} {|{\bf x}|}
- 
\frac 1 {|\Lambda|} \int_{\RealVect{2}} \frac {\exp(-|{\bf x}|/R)}
      {|{\bf x}|} d^2({\bf x})
\right)
\end{equation}
We evaluate this limit (omitting the factor $1/D_i$) in the appendix
as $-2.10671$ and denote its value as $C$.

Discarding the far piece in Equation~\eqref{eq:near-far}, assuming a
local hexagonal structure, and replacing the outer sum with an
integral on the right hand side of Equation~\eqref{eq:err-est1} gives
the following conjecture.

\begin{conjecture}\label{conj:simple}
Let $A$ be a compact $2$-manifold where $\mu^{1,A}$ is absolutely
continuous with respect to $\Hdr$, where the support of $\mu^{1,A}$ is
all of $A$, and where $U_1^{\mu^{1,A}}({\bf x}) = I_s(\mu^{s,A})$ for
all ${\bf x}\in A$.  Then
\begin{equation}\label{eq:second-term}
\lim_{N\to\infty}
\frac {{\mathcal E}_1({A},N) - I_1(\mu^{1,A})N^2 }
{N^{3/2}}
=
C
\sqrt{
\frac {\sqrt{3}} 2}
\int
\sqrt{
\frac 
{d\mu^{1,A}}
{d{\mathcal H}^2_A}
({\bf x})}
\,d\mu^{1,A}({\bf x}),
\end{equation}
where 
$$
C = 
\lim_{R\to \infty} \left(
\sum_{{\bf x} \in \Lambda \backslash \{0 \}} \frac {\exp(-|{\bf x}|/R)} {|{\bf x}|}
- 
\frac 1 {|\Lambda|} \int_{\RealVect{2}} \frac {\exp(-|{\bf x}|/R)}
      {|{\bf x}|} d^2({\bf x})
\right),
$$
and where $\Lambda$ is the unit hexagonal lattice.
\end{conjecture}

Conjecture~\ref{conj:simple} follows from a number of simplifying and
possibly unnecessary assumptions.  A broader conjecture that is closer
in form to Conjecture 2 given by Kuijlaars and
Saff~\cite{KuijlaarsSaff1} is
\begin{conjecture}\label{conj:KS-ext}
Let $A$ be a compact $2$-manifold, $0<s<2$, and $\mu^{s,A}$ 
absolutely continuous with respect to $\Hdr$, then 
$$
\lim_{N\to\infty}
\frac {{\mathcal E}_s({A},N) - I_s(\mu^{s,A})N^2 }
{N^{1 + s/2}}
=
C_s
\int
\sqrt{
\frac 
{d\mu^{s,A}}
{d{\mathcal H}^2_A}
({\bf x})}
\,d\mu^{s,A}({\bf x}),
$$
where 
$$
C_s = 6\left(\frac{\sqrt{3}}{8\pi}\right)^{s/2}\zeta(s/2)L_{-3}(s/2).
$$
\end{conjecture}
Here $\zeta$ is the analytic extension of the Riemann Zeta function and
$L_{-3}$ is the Dirichlet L-function given by
$$
L_{-3}(\alpha) = 1 - \frac 1 {2^\alpha} + \frac 1 {4^\alpha} - \frac 1
{5^\alpha} + \frac 1 {7^\alpha}  \cdots
$$ Conjectures~\ref{conj:simple} and~\ref{conj:KS-ext} both predict
$2.0 \times -0.553051$ for the coefficient of the $N^{3/2}$ term on
$\mathbb{S}^2$, and are in good agreement with energies on the sphere.

We now consider two additional numerical tests of these conjectures.
In the first test we shall look at the torus $\mathbb{T}^2$ using a
modest data set of low energy configurations.  However, we also need
an approximation of $\mu^{1,\mathbb{T}^2}$, and we turn to the work of
Brauchart, Hardin and Saff on sets of
revolution~\cite{BrauchartHardinSaff1}. In that work the authors begin
with the fact that for sets of revolution, the equilibrium measure
must be invariant under revolution. They develop a lower dimensional
minimization problem on the set, which when rotated, gives $A$.  While
the theory does not address the case $s=1$, we use their theory as a
recipe to approximate $\mu^{1,\mathbb{T}^2}$ numerically and present
the results in Table~\ref{table:energyCompare}.

\begin{figure}
\includegraphics[height=9cm]{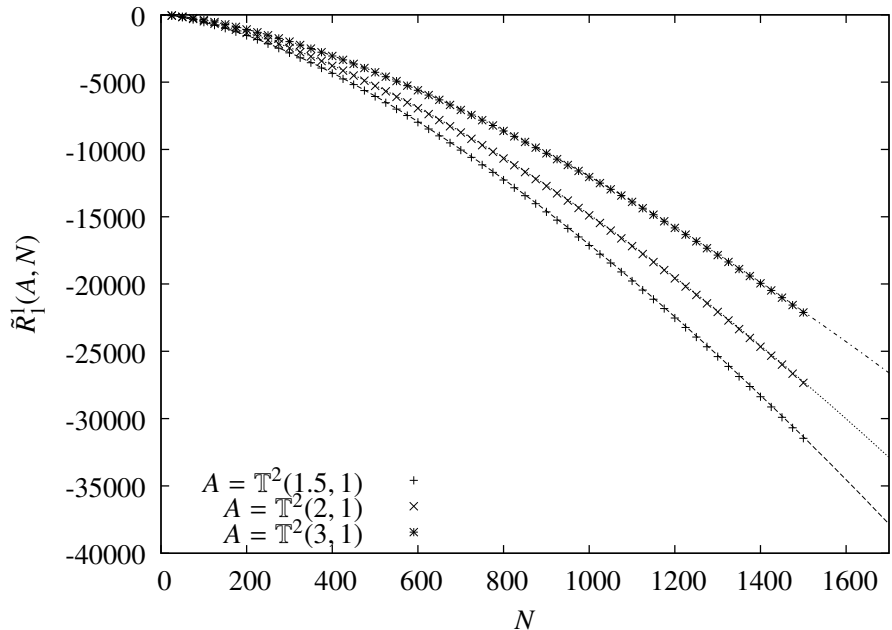}
\caption{\label{fig:SecondTerm} A plot of $\tilde R^1_1(A,N)$ for
  $A=\mathbb{T}^2(1.5,1)$, $A=\mathbb{T}^2(2,1)$ and $A =
  \mathbb{T}^2(3,1)$.  For each manifold, we've overlaid the
  prediction for the $N^{3/2}$ term given by
  Conjectures~\ref{conj:simple} and~\ref{conj:KS-ext}.}
\end{figure}

We denote the torus of major radius $l$ and minor radius $a$ by
$\mathbb{T}^2(l,a)$. Landkof~\cite[p. 166]{Landkof1} provides the
following formula for the energy of the equilibrium measure on the
torus:
\begin{equation}\label{eq:1-energy}
I_1\left(\mu^{1,\mathbb{T}^2(l,a)}\right) = 
\frac {2c}{\pi^2} \left[ \frac {Q_{-1/2}\left(\frac l a \right)}
  {P_{-1/2}\left(\frac l a \right)} + 2\sum_{n=1}^\infty \frac {Q_{n
      -1/2}\left(\frac l a \right)} {P_{n -1/2}\left(\frac l a
    \right)} \right],
\end{equation} where $c = \sqrt{l^2 - a^2}$ and where $P_\nu$ and $Q_{\nu}$ are
Legendre functions of the first and second kind. We use the GNU
Scientific Library~\cite{GSL} to evaluate the Legendre functions in
the above sum. In Table~\ref{table:energyCompare} we see good
agreement between the energies that result from extending the work
in~\cite{BrauchartHardinSaff1} to $s=1$ and the energies given
by~\eqref{eq:1-energy}.  Because the equilibrium measure is the unique
measure that minimizes the energy, we conclude that the measure
generated by applying the theory in~\cite{BrauchartHardinSaff1} to the
torus for $s=1$ generates a reasonable approximation of the
equilibrium measure on the torus.  Further, our numerical experiments
show that the support of the equilibrium measure is
$\mathbb{T}^2(l,a)$.  In Figure~\ref{fig:SecondTerm} we plot the
difference between the observed minimal energy and the first order
term, i.e. $\tilde R_1^1(A,N)$. We also plot the conjectured value for
the $N^{3/2}$ term using our numerical approximation of
$\mu^{1,\mathbb{T}^2}$.  The agreement suggests that
Conjectures~\ref{conj:simple} and~\ref{conj:KS-ext} appears to hold
for the torus.

\begin{table}
\centering
\begin{tabular}{ccccc}
$A$ & $M$ & Energy computed using~\cite{BrauchartHardinSaff1} & Energy
  computed with Equation~\eqref{eq:1-energy} & Relative error \\
\hline
$\mathbb{T}^2(1.5, 1)$ & $1000$ & $0.4782545$ & $0.47825526366953$ & $1.597\times
10^{-6}$ \\
$\mathbb{T}^2(2, 1)$ & $1000$ & $0.411239$ & $0.41123994225477$ & $2.291\times
10^{-6}$ \\
$\mathbb{T}^2(3, 1)$ & $950$ & $0.3234383$ & $0.323438867490233$ & $1.754\times
10^{-6}$ \\
\end{tabular}
\caption{\label{table:energyCompare} A comparison of the $s=1$ energy
  of the equilibrium energy computed in two ways on three different
  tori.  The first method uses the work of Brauchart, Hardin and
  Saff~\cite{BrauchartHardinSaff1} as a recipe for approximating the
  $s=1$ equilibrium measure. The second method uses
  Equation~\eqref{eq:1-energy}.  $M$ is the dimension of the
  discretized problem arising from~\cite{BrauchartHardinSaff1}.}
\end{table}

\begin{figure}
\includegraphics[height=9cm]{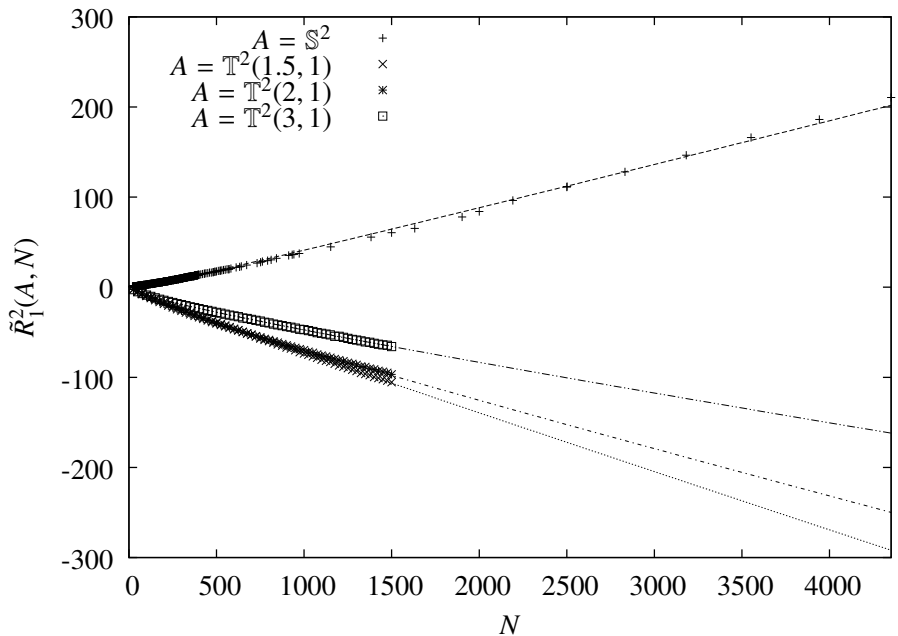}
\caption{\label{fig:ThirdTerm} A plot of $\tilde R^2_1(A,N)$ for $A =
  \mathbb{S}^2$, $A=\mathbb{T}^2(1.5,1)$, $A=\mathbb{T}^2(2,1)$ and $A
  = \mathbb{T}^2(3,1)$.  For each manifold we've overlaid the best fit
  of the form $\alpha N + \beta\sqrt{N}$.}
\end{figure}

We do not have a model beyond the second term.  However, our data
suggest the form of higher order terms.  In Figure~\ref{fig:ThirdTerm}
we've plotted the difference between the observed lowest energy and
the first two terms obtained from the transfinite diameter argument
and Conjecture~\ref{conj:simple}, i.e. $\tilde R^2_1(A,N)$.  We see
strong evidence that the third term is linear.  We fit $\tilde
R^2_1(A,N)$ to $\alpha N + \beta\sqrt{N}$ and report the values of
$\alpha$ and $\beta$ in Table~\ref{table:fitParams}.  To assign a
goodness of fit we would need to be able to estimate the error in our
estimates for the minimal energy.  However, useful estimates of such
errors from above are at least as hard as the formidable task of
bounding from below the minimal energy.

\begin{table}
\centering
\begin{tabular}{ccc}
$A$ & $\alpha$ & $\beta$ \\
\hline
$\mathbb{S}^2$ & 0.05123  & -0.3207 \\
$\mathbb{T}^2(1.5, 1)$ & -0.0616 & -0.3633 \\
$\mathbb{T}^2(2, 1)$ & -0.0462 & -0.7379 \\
$\mathbb{T}^2(3, 1)$ & -0.02780 & -0.6208 \\
\end{tabular}
\caption{\label{table:fitParams} Parameters from a best fit
  of $\alpha N + \beta \sqrt{N}$ to $\tilde R^2_1(A,N)$.}
\end{table}

\begin{figure}
\includegraphics[height=9cm]{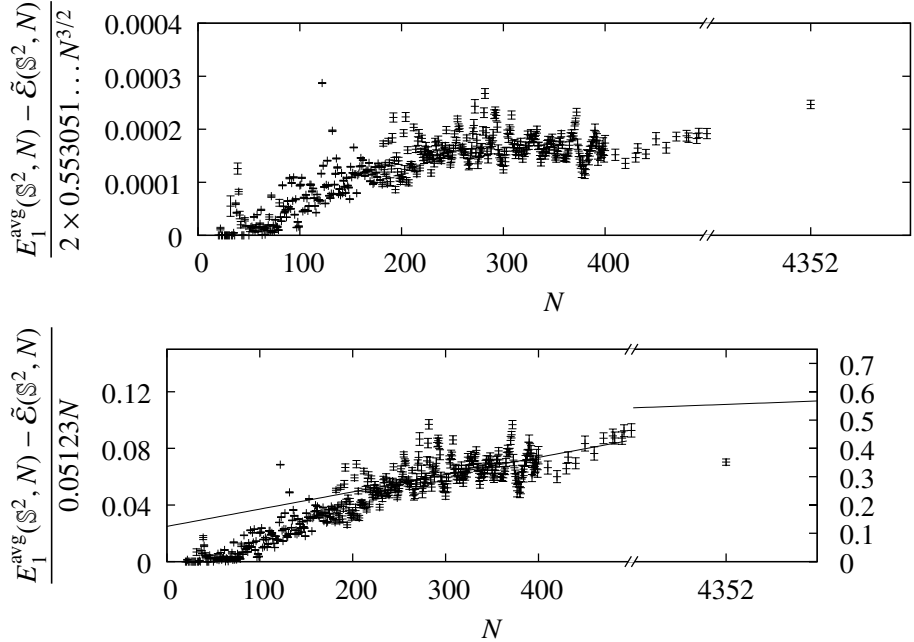}
\caption{\label{fig:variation_s1} The top plot shows the difference
  between the average energy of stable configurations and the minimal
  observed energy divided by the conjectured $N^{3/2}$ term.  In both
  plots the $x$ axis is broken to effectively display the data point
  at $N=4352$.  The bottom plot shows the same energy difference
  divided by the empirically obtained linear third term.  We have
  rescaled the right section of the lower plot and included a single
  line, plotted in both scales as reference.  The error bars in this
  plot, and following plots of this type, are the standard error of
  the mean of the energy of the stable configurations.}
\end{figure}

Conjectures~\ref{conj:simple} and~\ref{conj:KS-ext} are expressed in
terms of an integral over the equilibrium measure and a coefficient
derived from a sum over a hexagonal lattice.  The formulation of these
conjectures does not make any assumption about the location or
structure of the defects.  This would imply that, if stable
configurations differ from the minimal configuration only in the
structure and location of defects, then Conjectures~\ref{conj:simple}
and~\ref{conj:KS-ext} should approximate the average stable energy as
well.  This is our second test of the conjectures.  In the top of
Figure~\ref{fig:variation_s1} we see that the difference between the
average energy of stable configurations and the lowest observed energy
is bounded by three ten-thousandths of the conjectured $N^{3/2}$
term. In the bottom of Figure~\ref{fig:variation_s1} we see that this
difference between the average and minimal energies is substantially
larger when compared to the empirically obtained linear term
($.05123N$) for the minimal energy.  Indeed for our data at $N=4352$
the average and minimal energy differ by 30\% of the linear term.

The conclusion is that the first and second terms given by the
transfinite diameter and the conjectured $N^{3/2}$ term will predict
energies of stable and minimal configurations well, but the empirically
obtained linear third term reflects properties of the minimal
configuration that are absent in the stable configurations.  We assume
that these properties are the location and structure of the defects.

\subsection{The $s=0$ Case}

The problem of minimizing the $s=0$ energy is equivalent to the
problem of maximizing the product of pairwise distances of points, and
has received considerable attention from the mathematics community.
The seventh of Smale's eighteen problems for the twenty first
century~\cite{Smale1} is to develop an algorithm that will generate
rapidly a configuration, $\omega_N^*$, that satisfies $E_0(\omega_N^*)
- {\mathcal E}_0(\mathbb{S}^2,N) < C\log N$ for some constant $C$ that
does not depend on $N$.

One challenge in solving this problem is estimating ${\mathcal
  E}_0(\mathbb{S}^2,N)$ to at least ${\mathcal O}(\log N)$. Rakhmanov,
Saff and Zhou made progress in this direction by bounding the linear
term~\cite[Theorems 3.1 and 3.2]{RakSaffZhou2} by defining $C_N$ as
\begin{equation}\label{eq:log-second-term}
{\mathcal E}_0(\mathbb{S}^2,N) = 
-\frac 1 2 \log\left( \frac 4 e \right)N^2
-\frac 1 2 N \log N
+ C_N N,
\end{equation}
and showing
$$
-.225537540\ldots \le \liminf_{N\to\infty} C_N  
\quad\text{and}\quad
\limsup_{N\to\infty}C_N \le -.04699460\ldots
$$
In the same paper, those authors conjecture that 
\begin{equation}\label{eq:logConject}
{\mathcal E}_0(\mathbb{S}^2,N) = 
-\frac 1 2 \log\left( \frac 4 e \right)N^2
-\frac 1 2 N \log N
+ \alpha N 
+ \beta \log N 
+ {\mathcal O}(1).
\end{equation}

We fit 
$$
-\frac 1 2 \log\left( \frac 4 e \right)N^2
-\frac 1 2 N \log N
+ \alpha N 
+ \beta \log N 
+ \gamma
$$ to our minimal energies and find a best fit for $\alpha = -0.0547$,
$\beta=.6000 $ and $\gamma=-2.680 $.  The value of $\alpha$ we obtain
is in reasonable agreement with the value of $-0.052844$ obtained
empirically by Brauchart, Hardin and
Saff~\cite{BrauchartHardinSaff:2011}, and in stronger agreement with
the value of $-0.055605\ldots$ given in Conjecture
4~\cite{BrauchartHardinSaff:2011}.

\begin{figure}
\includegraphics[height=9cm]{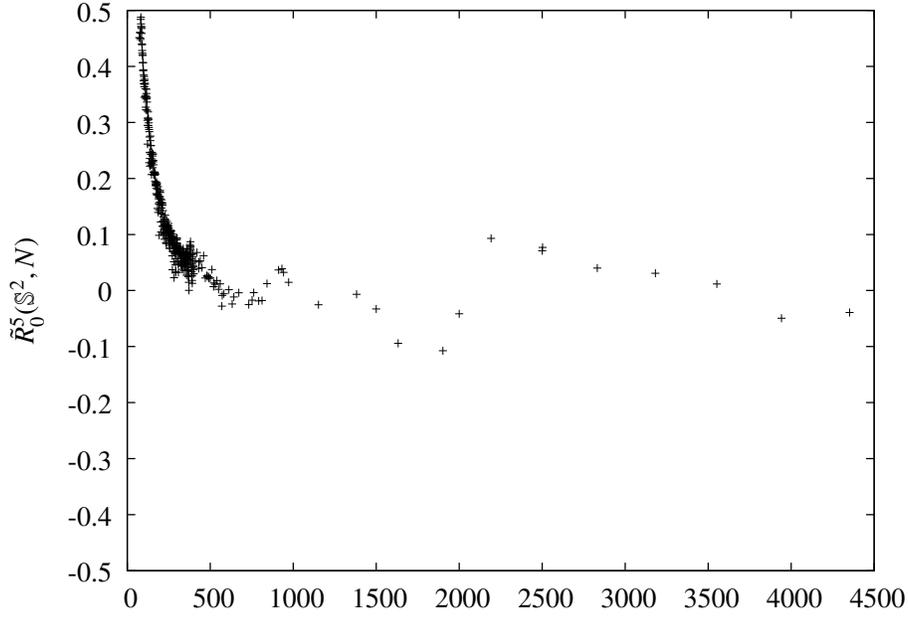}
\caption{\label{fig:s0} This is the observed minimal logarithmic
  energy minus a five term asymptotic expansion for the minimal
  energy.  We see evidence of a term that decreases with $N$.}
\end{figure}

We fit over a range of $N=501,\ldots,4352$ because the data with which
we have to work has behavior for $N\le 500$ that is not captured in
Equation~\eqref{eq:logConject}.  We plot the difference of the
observed lowest energy and the five term asymptotic expansion in
Figure~\ref{fig:s0}. It is worth noting that, for $N>500$, the
magnitude of this five term residual is less than $.2$ while the value
of ${\mathcal E}_0(\mathbb{S}^2,4352)$ is about $-3.6$ million.

\begin{figure}
\includegraphics[height=9cm]{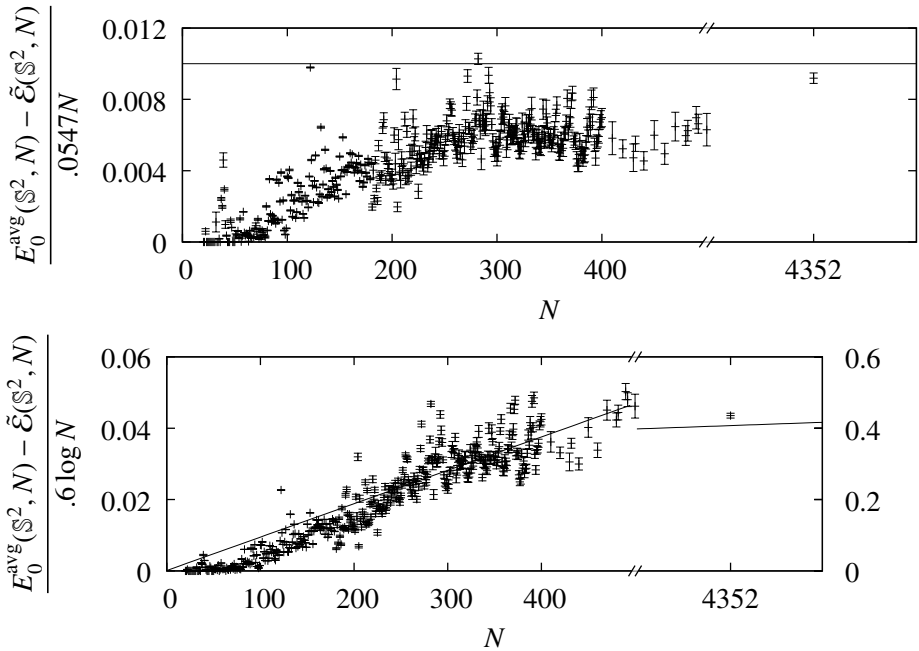}
\caption{\label{fig:variation_s0} The top plot shows the difference of
  the average and lowest observed $s=0$ energies divided by the linear
  term in an asymptotic expansion.  The bottom plots shows the same
  difference divided by the logarithmic term.}
\end{figure}

In Figure~\ref{fig:variation_s0} we compare the difference between the
average and minimal observed energies with the terms in the asymptotic
expansion.  For the data available, this energy difference is bounded
by about one percent of the empirically obtained linear term, as is
shown in the top plot.  That is, the difference between the average
energy of the stable configurations and the minimal observed energy is
growing roughly as $N/2000$.  It is worth comparing this with Figure 2
of~\cite{RakSaffZhou2} where the energy of constructively generated
spiral point configurations differs from an estimate of the minimal
energy by roughly $N/500$.  

The qualitative interpretation that the data in the upper plot in
Figure~\ref{fig:variation_s0} are bounded while the data in the lower
plot are growing implies that the first three terms in the asymptotic
expansion describe the energy of stable configurations as well as the
energy of minimal configurations, while the logarithmic term in the
asymptotic expansion will reflect properties of the minimal
configurations that are absent in most stable configurations.  This
implies that solving Smale's seventh problem will require some
understanding of the defects.

\subsection{The $s=2$ Case}

\begin{figure}
\includegraphics[height=9cm]{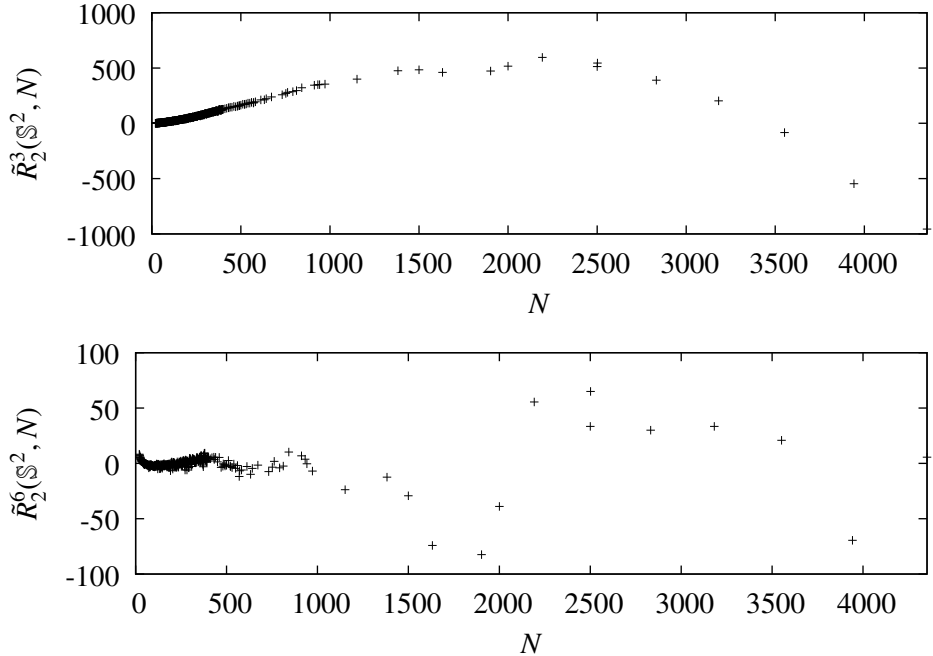}
\caption{\label{fig:s2} The top plot shows the residual after the
  three term expansion given by Conjecture 5
  from~\cite{BrauchartHardinSaff:2011}.  The bottom plot shows the
  residual with three additional terms. }
\end{figure}

The Riesz kernel $k_2$ is not locally integrable on a $2$-manifold and
the potential theoretic arguments cannot provide a first order term.
Initial results for the leading order term on the sphere are given by
Kuijlaars and Saff~\cite[Theorem 3]{KuijlaarsSaff1}.  These results
were generalized to a class of sets that include $C^1$ manifolds by
Hardin and Saff~\cite[Theorem 2.4]{HardinSaff1}.  Combining these
results with Conjecture 5 from~\cite{BrauchartHardinSaff:2011}, one
has an asymptotic expansion of the form
$$
{\mathcal E}_2(\mathbb{S}^2, N) = \frac 1 4 N^2 \log N + \alpha N^2 + {\mathcal O}(1)
$$ The conjectured value for $\alpha$ is
$-0.08576841030090248365\ldots$

We fit the available data to 
$$
\frac 1 4 N^2 \log N + \alpha N^2 + \varepsilon
$$ and find that $\alpha = -0.085079$.  However, the difference
between the observed minimal energies and the best fit, shown in the
top of Figure~\ref{fig:s2}, has considerable structure.  One
hypothesis is that the form of the expression used for the fit is
not correct.  Making the arbitrary decision to include the same
sequence of terms found in the expansion for the logarithmic energy,
we fit
$$
\frac 1 4 N^2 \log N + \alpha N^2 + \beta N\log N + \gamma N + \delta
\log N + \varepsilon
$$ to our data, and when we fit the above, we found $\alpha =
-0.085417$ and $\beta = .4415$.  The residuals associated with the
best fit of this augmented asymptotic expansion is shown in the lower
plot of Figure~\ref{fig:s2}.

\begin{figure}
\includegraphics[height=9cm]{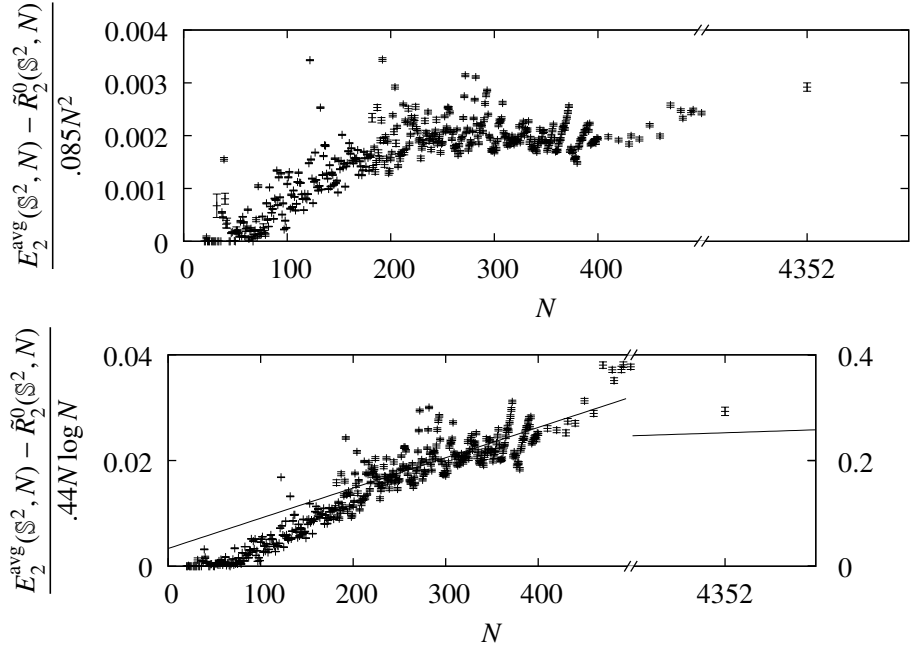}
\caption{\label{fig:variation_s2} These two plots are the ratios of
  the difference between the average energy of stable configurations
  and the minimal observed energy to two terms in the asymptotic
  expansion for the energy.  The top plot shows this difference
  compared to the empirically obtained $N^2$ and the bottom plot
  compares this difference with the empirically obtained $N\log N$
  term.}
\end{figure}

Figure~\ref{fig:variation_s2} shows the growth of the difference
between the average energy of stable configurations with the minimal
observed energy divided by the $N^2$ term in the top plot and an
empirically obtained $N\log N$ term in the bottom plot.  If one accepts
that the data in the top plot is bounded, and the data in the bottom
plot is growing, then one would conclude that the first two terms in
the asymptotic expansion for the $s=2$ energy describe the energy of
stable configurations as well as the minimal energy to about three
parts in one thousand, while the next term, possibly an $N\log N$
term, would reflect properties of the minimal configurations absent in
most stable configurations.

\subsection{The $s=3$ Case}

\begin{figure}
\includegraphics[height=9cm]{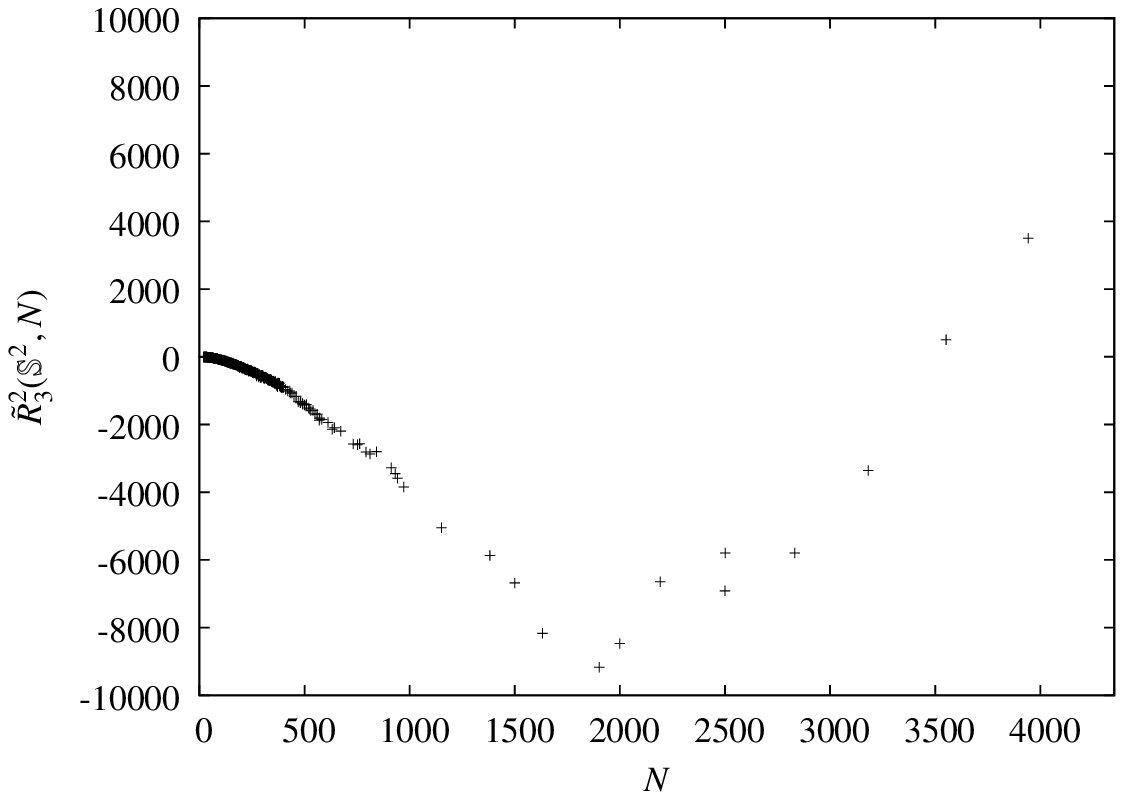}
\caption{\label{fig:s3} The difference between the observed minimal
  $s=3$ energy and the two term expansion for the $s=3$ energy.}
\end{figure}

The Riesz kernel $k_3$, like $k_2$, is not locally integrable on
$2$-manifolds.  Early progress toward the leading order term for the
asymptotic expansion of minimal $N$-point energy on the
sphere~\cite[Theorem 2]{KuijlaarsSaff1} shows that, if the leading
order term exists for any $s>d$, the leading order term has the form
$N^{1+s/2}$.  Kuijlaars and Saff further conjecture that
\begin{equation}\label{eq:limitExist}
\lim_{N\to\infty} \frac {{\mathcal E}_s(\mathbb{S}^2,N)}{N^{1+s/2}} = 
\left(
\frac {\sqrt{3}}{8\pi}
\right)^{s/2}
\zeta_\Lambda(s) =: \alpha
\end{equation}
where $\Lambda$ is again the hexagonal lattice and $\zeta_\Lambda$ is
the associated zeta function -- the sum of the reciprocals of the
non-zero distances in $\Lambda$ raised to the argument.  The existence
of the limit in~\eqref{eq:limitExist}, and hence the first order term,
was established for a broad class of sets by Hardin and
Saff~\cite{HardinSaff1} and strengthened by Borodachov, Hardin and
Saff~\cite{BHS1}, although the value of the limit has still not been
proven.  The natural assumption of a local hexagonal structure is
implicit in the conjecture as $\Lambda$ is the hexagonal lattice.  We
compute this leading term, via the factorization
presented~\cite{KuijlaarsSaff1} to get a value of $\alpha = 2.0\times
0.0998139\ldots$.  The second order term is
conjectured~\cite[Conjecture 3]{BrauchartHardinSaff:2011} to be $\beta
N^2$ where $\beta$ is given as the analytic extension, in $s\in
\mathbb{C}$, of $I_s(\mu^{s,\mathbb{S}^2})$ to the case
$s=3$. Following~\cite[Equation 10]{BrauchartHardinSaff:2011} we
compute the coefficient as $\beta=-.25$.

Fitting the expression 
\begin{equation}\label{eq:tofit}
\alpha N^{1+3/2} + \beta N^2, 
\end{equation} with $\alpha$ fixed at the value given in~\eqref{eq:limitExist}, to
our data for $N=20,\ldots,4352$ gives a value of $\beta =
-0.22\ldots$.  The addition of terms of the form $\gamma N^{1.5} +
\delta N + \varepsilon N^{.5}$ does not substantially change the value
for $\beta$ obtained through such a fitting procedure.  If we fit
Expression~\eqref{eq:tofit} to the data and let $\alpha$ vary we
obtain $\alpha = 2.0\times 0.099878$ and $\beta = -0.2349\ldots$.

The difference between the observed lowest energy and the fit, shown
in Figure~\ref{fig:s3} shows considerable structure, suggesting that
either the form to which we fit is not correct, or that the energies
with which are working are not minimal.

\begin{figure}
\includegraphics[height=9cm]{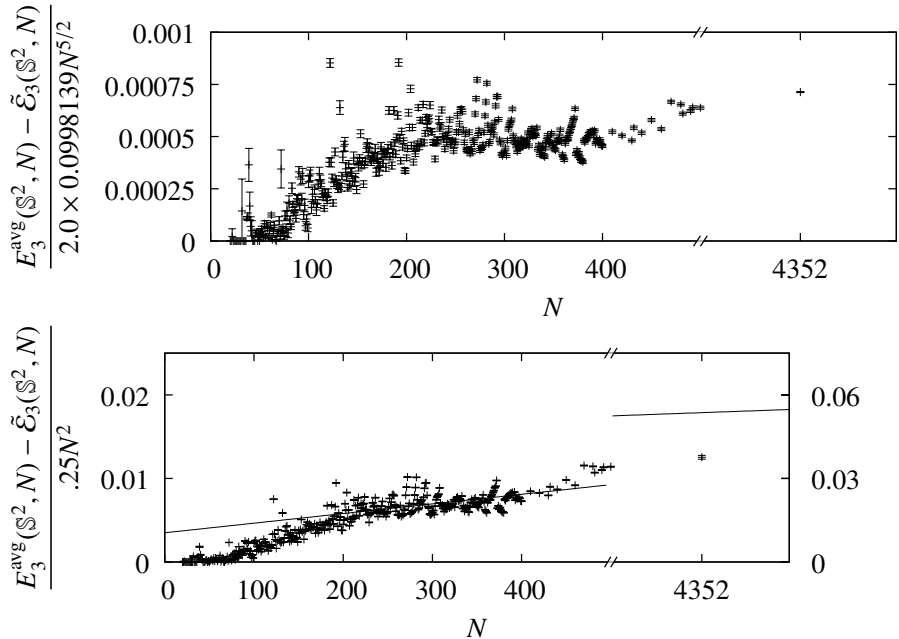}
\caption{\label{fig:variation_s3} Here we present, in the top plot,
  the difference between the average of the $s=3$ energies of stable
  configurations and the minimal observed $s=3$ energy divided by the
  leading order ($N^{5/2}$) term estimate.  The bottom plot shows the
  difference between the highest and lowest observed energies divided
  by the difference between the observed and conjectured second order
  term.}
\end{figure}

We plot the difference between the average and minimal energies in
Figure~\ref{fig:variation_s3}.  The upper plot suggests that this
difference is small compared to the leading order term.  The lower
plot compares this difference to the conjectured second order term.
This difference is about $4$ percent of the conjectured second order
term at $N=4352$.  However, the difference between the empirically
obtained coefficient for the second order term and the conjectured
coefficient is $12$ percent of the conjectured second order term.  If
our measurement of the second order coefficient differs from the
conjectured value because our lowest observed energies are not the
minimal energies, then the minimal energies differ from the lowest
observed energies by several times the difference between the average
and minimal energies.

\section{Conclusions}\label{sec:conclusion}

We've used numerically generated candidates for $s$-energy minimizing
configurations to assess conjectures for higher order terms in
asymptotic expansions for the minimal $s$-energy.  In addition we've
developed a large library of stable configurations and compared the
average of the energies of the stable configurations with the energies
of the candidate minimal configurations to approximate a lower bound
on the difference between the average and minimal energy.  

\subsection{Comparison of conjecture and numerical experiment} 

For $s=1$ we find that existing conjectures for the second order term
on the sphere appear appear to hold when extended to the torus, and
that the third term appears to be linear.  For the sphere a
straightforward fit suggests a value of $0.0513$ as the coefficient of
this linear term.  For $s=0$ the conjectured forms for the asymptotic
expansion gave rise to an expression that agreed, for $N>500$, with
our observed minimal energies to one part in thirty million.  Using a
fit for the linear term gives a value of $-0.0547$, while the
conjectured value is $-0.055605\ldots$.  For $s=2$ the conjectured
form of the asymptotic expansion left considerable structure,
suggesting that either the form of the fit was wrong or that the
energies with which we had to work were not minimal.  Two fits,
assuming different forms of the asymptotic expansion, gave values for
the coefficient of the conjectured second order $N^2$ term of
$-0.085079$ and $-0.085417$.  The conjectured value is
$-0.085768\ldots$ For $s=3$, the conjectured coefficient of the first
order term is $2.0\times 0.0998139\ldots$, while fitting our data
gives $2.0\times 0.099856\ldots$.  The second order term is
conjectured to be $-.25 N^2$.  Fitting our data suggests a coefficient
of $-.22$.

\subsection{Identification of terms that likely reflect defect structure}

For $s=1$ the difference between the average and lowest observed
energy was small compared to the $N^{3/2}$ term, and appeared to be
growing compared to an empirically obtained linear term.  For the
$s=0$ case this difference appeared to be bounded when compared to the
linear term, but growing when compared to the $\log N$ term.  This
suggests that an arbitrary sequence of stable configurations will not
be a solution to Smale's seventh problem.  For $s=2$ this difference
was small compared to the $N^2$ term, but growing compared to $N\log
N$.  For $s=3$ this difference was small compared to the leading order
term.

Because the stable configurations differ from minimal configurations
in the location and structure of defects, we infer that the energy
difference between stable states and minimal configurations is the
energy scale at which defects play a role.  And that theoretical
models for the terms identified above will require an understanding of
the role of defects.

\appendix
\section{Computing the limit in \eqref{eq:to-evaluate}}

We want to compute
$$
\lim_{R\to\infty} \left(
\sum_{{\bf x} \in \Lambda \backslash \{0\}} \frac 1 {|{\bf x}|}
e^{-\frac {|{\bf x}| } {R } }
- \frac 1 {|\Lambda|} \int_{\mathbb{R}^2} \frac 1 {|{\bf x}|} 
e^{-\frac {|{\bf x}| } {R } }d^2{\bf x}
\right),
$$
where $d^2{\bf x}$ indicates integration with respect to area.  For
convenience we let
$$
P_R({\bf x}) := \frac 1 {|{\bf x}|}
e^{-\frac {|{\bf x}| } {R } }.
$$
We have
$$
\sum_{{\bf x} \in \Lambda \backslash \{0\}} \frac 1 {|{\bf x}|}
e^{-\frac {|{\bf x}| } {R } }
= \sum_{{\bf x} \in \Lambda \backslash \{0\}} P_R({\bf x}) 
e^{-|{\bf x}|} + 
\left(\sum_{{\bf x} \in \Lambda} P_R({\bf x}) 
\left( 1 - e^{-|{\bf x}|} \right) \right)
- 
P_R({\bf 0}) 
\left( 1 - e^{-|{\bf 0}|} \right) .
$$ We interpret $P_R({\bf 0}) \left( 1 - e^{-|{\bf 0}|} \right)$ as
the limit as ${\bf x} \to 0$ of the function $f({\bf x}) = P_R({\bf
  x}) \left( 1 - e^{-|{\bf x}|} \right)$.  Applying the Poisson
Summation formula gives
$$
\sum_{{\bf x} \in \Lambda} P_R({\bf x}) 
\left( 1 - e^{-|{\bf x}|} \right) 
= 
\frac 1 {|\Lambda|} \sum_{\xi \in \Lambda^* \backslash \{0\}} \left(P_R({\cdot}) 
\left( 1 - e^{-|{\cdot}|} \right)\right)\hat\,(\xi)
+ \frac 1 {|\Lambda|} \hat{P_R}(0) - 
\frac 1 {|\Lambda|} \left(P_R(\cdot)e^{-|\cdot|}\right)\hat\,(0)
$$
For some $\alpha$ we compute $\hat P_{\alpha}$ as 
$$
\hat P_{\alpha} (\xi) = 
\int_{\mathbb{R}^2} e^{- 2\pi i \xi\cdot{\bf x}} 
\frac 1 {|{\bf x}|}
e^{-\frac {|{\bf x}| } {\alpha } } d^2{\bf x}.
$$ Both $P_{\alpha}$ and $\hat P_{\alpha}$ are rotationally symmetric,
so we can choose $\xi = (0,1)|\xi|$ and integrate in polar coordinates
-- this change to polar coordinates leads to a convenient cancellation
when $s=1$ -- to get
$$
\hat P_{\alpha} (\xi) = 
\int_0^\infty
e^{-\frac r \alpha}
2\pi \frac 1 {2 \pi}
\int_0^{2\pi}
e^{-i (2\pi|\xi|r)\sin \theta} d\theta\, dr
= 2\pi 
\int_0^\infty e^{ - \frac 1 \alpha r} J_0 ( 2\pi |\xi| r ) dr.
$$
Recognizing the right most integral as the Laplace Transform of the
Bessel Function $J_0$ gives
$$
\hat P_\alpha (\xi)= 
\frac { 2\pi } {\sqrt{\left(\frac 1 \alpha\right)^2 + (2 \pi |\xi|)^2}}.
$$
Note also that $P_Re^{-|\cdot|} = P_{\frac R {1 + R}}$ and that 
$$
\hat P_R (0) = 
\frac 1 {|\Lambda|} \int_{\mathbb{R}^2} \frac 1 {|{\bf x}|} 
e^{-\frac {|{\bf x}| } {R } }d^2{\bf x},
$$
which allows us to collect terms and write the quantity we would like to
compute as the limit as $R\to\infty$ of
$$
\sum_{{\bf x} \in \Lambda \backslash \{0\}} \frac 1 {|{\bf x}|}
e^{-\frac {|{\bf x}| } {R } }
- \frac 1 {|\Lambda|} \int_{\mathbb{R}^2} \frac 1 {|{\bf x}|} 
e^{-\frac {|{\bf x}| } {R } }d^2{\bf x}
=
$$
\begin{eqnarray*}
\sum_{{\bf x} \in \Lambda \backslash \{0\}} P_R({\bf x}) 
e^{-|{\bf x}|} \\
+\frac 1 {|\Lambda|} \sum_{\xi \in \Lambda^* \backslash \{0\}} 
\left(\hat P_R(\xi) -
\hat P_{\frac R {1 + R}}(\xi)\right) \\
- \frac 1 {|\Lambda|} \hat P_{\frac R { 1+ R}}( 0 ) \\
- \left. P_R({\bf x}) 
\left( 1 - e^{-|{\bf x}|} \right) \right|_{{\bf x} = 0}.
\end{eqnarray*}
The limit is well defined for each term. For the first term we have
$$
\lim_{R\to \infty} 
\sum_{{\bf x} \in \Lambda \backslash \{0\}} P_R({\bf x}) e^{-|{\bf x}|} 
= 
\sum_{{\bf x} \in \Lambda \backslash \{0\}} \frac 1 {|{\bf x}|} e^{-|{\bf x}|}, 
$$
by monotone convergence. For the second term we have
$$
\lim_{R\to\infty}
\frac 1 {|\Lambda|} \sum_{\xi \in \Lambda^* \backslash \{0\}} 
\left(\hat P_R(\xi) -
\hat P_{\frac R {1 + R}}(\xi)\right) 
$$
\begin{eqnarray*}
=&\displaystyle{
\lim_{R\to\infty}
\frac  { 2\pi } {|\Lambda|} \sum_{\xi \in \Lambda^* \backslash \{0\}} 
\left(
\frac 1 {\sqrt{\left(\frac 1 R\right)^2 + (2 \pi |\xi|)^2}} - 
\frac 1 {\sqrt{\left(\frac {1+R} R\right)^2 + (2 \pi |\xi|)^2}}.
\right) }\\
=&
\displaystyle{\lim_{R\to\infty}
\frac  { 2\pi } {|\Lambda|} \sum_{\xi \in \Lambda^* \backslash \{0\}} 
\left(
\frac {{\left(\frac {1+R} R\right)^2 } - {\left(\frac 1 R\right)^2 }}
{
\sqrt{\left({\left(\frac 1 R\right)^2 + (2 \pi |\xi|)^2}\right)\left
({\left(\frac {1+R} R\right)^2 + (2 \pi |\xi|)^2}\right)}\left(
\sqrt{{\left(\frac 1 R\right)^2 + (2 \pi |\xi|)^2}} + 
\sqrt{{\left(\frac {1+R} R\right)^2 + (2 \pi |\xi|)^2}}
\right)
}
\right) }\\
=&
\displaystyle{\frac  { 2\pi } {|\Lambda|} \sum_{\xi \in \Lambda^* \backslash \{0\}} 
\left(
\frac 1
{
2 \pi |\xi| \sqrt{1 + (2 \pi |\xi|)^2}
\left(
2 \pi |\xi| + 
\sqrt{1 + (2 \pi |\xi|)^2}
\right)
}
\right) },
\end{eqnarray*}
by dominated convergence. By direct evaluation, the third and fourth
terms are
$$
-\frac 1 {|\Lambda|}\lim_{R\to\infty} 
\hat P_{\frac R { 1+ R}}( 0 ) = 
-\frac {2\pi} {|\Lambda|}
$$
and
$$
- \lim_{R\to\infty} \left. P_R({\bf x})\left( 1 - e^{-|{\bf
  x}|} \right) \right|_{{\bf x} = 0} = -1.
$$
We are left with 
$$
\sum_{{\bf x} \in \Lambda \backslash \{0\}} \frac 1 {|{\bf x}|} e^{-|{\bf x}|} 
+ \frac  { 2\pi } {|\Lambda|} \sum_{\xi \in \Lambda^* \backslash \{0\}} 
\left(
\frac 1
{
2 \pi |\xi| \sqrt{1 + (2 \pi |\xi|)^2}
\left(
2 \pi |\xi| + 
\sqrt{1 + (2 \pi |\xi|)^2}
\right)
}
\right) 
-\frac {2\pi} {|\Lambda|}
-1.
$$
We shall choose $\Lambda$ to be the hexagonal lattice, that is the
lattice generated by the vectors $(1,0)$ and $\left(\frac 1 2 , \frac
{\sqrt 3} 2 \right)$.  In this case $\Lambda^*$ is generated by the vectors
$\left( 0 , \frac 2 {\sqrt 3 } \right)$ and $\left( 1 , \frac 1 {\sqrt
3} \right)$. Finally $|\Lambda| = \frac {\sqrt 3} 2$.

\end{document}